# A Data-Driven Approach to Dynamically Adjust Resource Allocation for Compute Clusters

Francesco Pace[1], Dimitrios Milios[1], Damiano Carra[2], Daniele Venzano[1] and Pietro Michiardi[1]

[1]*Data Science Department, Eurecom, Biot Sophia-Antipolis, France*

[2]*Computer Science Department, University of Verona, Verona, Italy*

[1] {*pace,milios,venzano,michiard*}*@eurecom.fr*   [2]*damiano.carra@univr.it*

## Abstract

Nowadays, data-centers are largely under-utilized because resource allocation is based on reservation mechanisms which ignore actual resource utilization. Indeed, it is common to reserve resources for peak demand, which may occur only for a small portion of the application life time. As a consequence, cluster resources often go under-utilized.

In this work, we propose a mechanism that improves cluster utilization, thus decreasing the average turnaround time, while preventing application failures due to contention in accessing finite resources such as RAM. Our approach monitors resource utilization and employs a data-driven approach to resource demand forecasting, featuring quantification of uncertainty in the predictions. Using demand forecast and its confidence, our mechanism modulates cluster resources assigned to running applications, and reduces the turnaround time by more than one order of magnitude while keeping application failures under control. Thus, tenants enjoy a responsive system and providers benefit from an efficient cluster utilization.

## 1 Introduction

Data-center efficiency is a subject that attracted a vast amount of research [6, 65, 49, 62, 54, 11, 2]. Recently, the cloud computing paradigm, both in its public and private forms, fueled the proliferation of a wide array of resource management tools [62, 54, 17, 31] aiming at an efficient operating point, where cluster resources are fully utilized. Despite such efforts, data-center resources go often under utilized, as shown in recent traces from large-scale production clusters [53, 63]: in most cases ($\sim 80\%$) resource utilization is less than 40% or 80% of the allocated resources depending on application types.

Current approaches that address efficiency requirements fall in two broad categories. The first involves methodologies that steer tenants' behavior through the design of incentive mechanisms; tenants are endowed with the task of optimizing their cost to operate their applications, whereas providers operate on prices to regulate the allocation of idle resources. Such approaches are largely adopted by public cloud providers [6]. The second category concerns approaches that operate at the system level, and propose mechanisms that allocate resources based on tenants' reservations [23, 49, 31, 62, 54, 17, 5]. Essentially, existing approaches either let tenants reason in terms of value and costs [6], or let the system determine how to avoid wasting scarce and costly resources.

In this paper, we discuss a methodology that belongs to the second category: we present a mechanism that dynamically adjusts resources allocated to running applications according to their expected utilization, as opposed to a static allocation based on tenants' reservations. In the context we consider, we define as *applications* the use of distributed frameworks such as Apache Spark [4] and Google TensorFlow [27] that include different components to produce work.

**Reservation centric resource allocation.** In most private or public cloud systems, users gain access to computing resources by specifying the amount of resources required to run their application, in the form of a reservation request. Upon receiving a request, the cluster **scheduler** decides which application to serve based on the scheduling policy the provider implements (e.g., First-In-First-Out (FIFO)). Cluster schedulers operate according to several variants of objective functions, including fairness across users, service-level objectives, and various measures of performance. In this work, we focus on two common optimization objectives: *(i)* average turnaround time (also called completion time) and *(ii)* cluster utilization [54, 37, 3]. The first metric accounts for the average time requests spend in the system (queuing and execution times). The second metric considers the utilization of the available resources. Optimizing for such objectives translates in high system responsiveness, which is desirable for both tenants and providers.

Cluster schedulers use mechanisms to provision and manage resources: given a **resource request**, the resource manager determines its admission in the cluster based on **reservation** information.[1] An admitted request triggers a **resource allocation** procedure, which concludes with reserved resources being allocated to the request [54]. In most system implementations, the concept of reservation and allocation coincide, although neither is representative of the true **resource utilization** a request might induce on the system. In fact, resource utilization is generally not constant throughout a request lifetime, and fluctuates according to application behavior [64].

The main consequence for current cloud environments is that reservation requests are engineered to cope with **peak** resource demands of an application, which is one key factor that induces poor system utilization, and ultimately, negatively impacts system efficiency. This is exacerbated by coarse-grained reservation specifications: instance *flavors* exhibit discrete gaps in terms of resource units. In fact, picking the right configuration for cloud applications (and in particular for the "big data" applications we consider in this paper) is a daunting task [1], which requires sophisticated optimization mechanisms going beyond human tuning abilities.

Thus, mechanisms to reduce **resource slack**, which is defined as the difference between resource allocation and utilization, are truly needed, for they can prevent clusters from denying admission to new requests which would queue up, while spare capacity goes unused.

**Problem Statement.** We study the problem of cluster efficiency by reducing the resource slack induced by reservation-centric application schedulers, which match allocation to reservation. To do so, we introduce a new mechanism that **predicts** the resource utilization and adjusts the resource allocation accordingly. The main challenge to face is that prediction errors may have problematic consequences, since sudden spikes could wreak havoc the system [62]. When dealing with finite resources such as RAM, in fact, not providing the correct amount of resources leads to application failures. Careful engineering would suggest to introduce a buffer that will act as "safe-guard" to prediction errors. This results in a **trade-off**, since on the one hand the safe-guard buffer should be small to minimize slack, while on the other hand it should be sufficiently large to prevent application failures.

Previous works (a detailed description is provided in Section 2) usually consider shareable resources, such as CPU, where the effect of wrong resource dimensioning does not translate into application failures. Other approaches consider resource over-provisioning, where the slack is not continuously optimized, and where the application failures can be unpredictable and are taken care by the Operating System (OS).

In our approach, we leverage on three key ideas: prediction confidence, application elasticity and controlled failures. In the prediction process, most of the tools provide additional information about the **confidence** of the prediction. We use such information to dynamically adapt the safe-guard buffer that should prevent application failures. In addition, the frameworks, on which the applications are based, are composed by several elements that are characterized by either a **core** or **elastic** nature [42]. Core components are compulsory for a framework to produce useful work (e.g, Apache Spark requires a controller, a master, and one worker); elastic components, instead, optionally contribute to a job, e.g. by decreasing its runtime. An application that features only core components is called **rigid**, whereas applications with a mix of core and elastic components are called **elastic**. If the resource demand is higher than the available resources, we intervene (when possible) on elastic components to avoid application failures. As a last step, should the previous two mechanisms not be sufficient to provide enough resources, we explicitly decide which application should fail so that to minimize the amount of wasted work.

**Contributions.** In this paper we present our design of a data-driven resource shaping mechanism that improves cluster utilization, thus decreasing the average turnaround time, while preventing application failures due to resource contention. Our approach monitors resource utilization and relies on online forecasting of resource demand to modulate allocated resources such as they approximate utilization patterns well. Our experiments, that we conduct on a system simulator as well as a full-fledged implementation using real-life data-center traces, indicate substantial gains over existing alternatives. In summary, the contributions we present in this work are as follows:

- We present the design of a mechanism that dynamically adjusts resources allocated to applications by an existing scheduler. In this work, we target a specific family of application schedulers, and materialize our ideas for such systems.

- We compare parametric and non-parametric machine learning methodologies for the forecasting of resource utilization. In particular, we focus on accurate quantification of uncertainty, which is used to steer system parameters to safeguard against unexpected resource demand peaks.

- We perform an extensive simulation campaign using publicly available production traces from Google data-centers, and discuss about the trade-off that an optimistic vs. a pessimistic approach to application preemption entails. We also present a full-fledged implementation of our mechanism, that we use in an academic compute cluster serving hundreds of students and researchers. Our results indicate substantial improvements in terms of efficiency, which translate in a system capable of ingesting a heavier workload with the same number of machines.

---

[1] In our prose, we neglect several important technical details that are however irrelevant to our point, such as quota management, security aspects, and concurrency control, to name a few.

The remainder of the paper is organized as follows. In Section 2 we review the related literature. In Section 3 we present our system design, and we validate our ideas using a simulation campaign in Section 4. We present our prototype implementation in Section 5 and its evaluation in Section 5.1. Finally we conclude in Section 6.

## 2   Related Work

Resource allocation has been approached in many different ways in the literature [62, 35, 34, 11, 2, 14, 13, 23, 28, 49, 38, 56, 30, 6, 65, 43, 16, 44, 15].

The authors in [35, 34] use feedback control loop which requires every *framework* to periodically send application-specific information to the scheduler, which is used to steer resource allocation. In contrast, our approach does not require such instrumentation, as it is application agnostic: we use general metrics to dynamically adjust resources allocated to running applications.

The authors in [11] introduce a reservation-based scheduler and propose a Reservation Definition Language (RDL) that allows users to declaratively reserve access to cluster resources. They formalize the planning of current and future cluster resources as a mixed-integer linear programming problem and they integrate their work in YARN [61]. In our work, we avoid delegating this task to users by asking them to specify such information; generally, users have no knowledge of how their applications will behave.

The authors in [43] develop a feedback control loop for virtual machines, using a simple regression model to forecast future allocation. They show that it is possible to reduce the CPU resource slack, but they do not address memory and the consequences that under-provisioning such resource has on applications, as we do in our work.

The authors in [7] adopt a distributed scheduling architecture, whereby each scheduler aims at minimizing task completion time by careful placement strategies that use estimates of task runtime and their resource utilization. Contrary to our work, they use over-provisioning of resources and they tackle conflicts in an optimistic-manner. Our approach cooperates with an existing scheduler, instead of replacing it, and does not use task runtime to adjust cluster resources allocated to applications.

Some other works [6, 56] propose to address the problem with economics principles. In particular, in [56] the authors build a pricing model that enables infrastructure providers to incentivize their tenants to use graceful degradation, a self-adaptation technique originally designed for constructing robust services that survive resource shortages. The authors in [6], present a framework for scheduling and pricing cloud resources, aimed at increasing the efficiency of cloud resources usage by allocating resources according to economic principles. However, they achieve that by allocating more capacity than what is physically available, i.e., over-provisioning, which is a solution prone to uncontrolled failures[2] when utilization exceeds available resources.

Finally, works such as [33, 40, 49, 38, 2, 14, 13, 23, 28], focus either on resource placement or on meeting Service Level Objective (SLO). In the first case they relate to a packing problem and try to optimize it; Karanasos et al [33] suggest to dynamically re-balance the load across hosts if the packing performed at a certain time leads to uneven loaded hosts. In the second case they leverage the elasticity of some frameworks and they increase resources for applications that are falling behind on their SLO. Our work is orthogonal to such methods and can leverage them to improve the system performance.

The authors in [66] propose task scheduling and data placement techniques that rely on historical resource utilization. Specifically, they process the history of CPU utilizations using the Fast Fourier Transform (FFT). Leveraging the $k$-Means algorithm, they cluster patterns in three categories: periodic, constant and unpredictable. They exploit the patterns of periodic and constant categories to improve the quality of task scheduling.

Albeit all these works are valid and propose their own vision of the problem, they share one element: although some of them address a multi-dimensional packing problem for provisioning resources to applications, when it comes to reclaiming resources granted to applications they mostly focus on "time sharable" resources, like the CPU, rather than "finite" resources like Memory.[3] As a consequence, such methods are limited to improve system efficiency from the perspective of CPU utilization.

An example of prior work that modulates "finite" resources is Borg [62]. Borg features a resource reclamation system that seizes unused resources and offers them to other applications. The authors study the impact of wrong memory reallocation on running tasks, which causes resource contention: the OS enters a special state to kill processes that are OOM. The authors present different levels of "rigidity" for their reclamation system (baseline, medium and aggressive) and show both the benefit and the number of OOMs events for each of them. They conclude by accepting the trade-off obtained by the medium setting. Instead, we present a dynamic allocation system that relies on online resource forecasting, with accurate quantification of uncertainty. In addition, we seek to gain control over the OS and minimize application failures events while maximizing the resource utilization.

---

[2]The OS kills processes due to Out Of Memory (OOM) following its own algorithm.

[3]On the one hand, a resource is considered "time sharable" when the OS is able to use time sharing for scheduling it, and thus it does not impose limits on its availability. On the other hand, "finite" resources are those that cannot be sliced in time and thus cannot be effectively shared by multiple processes.

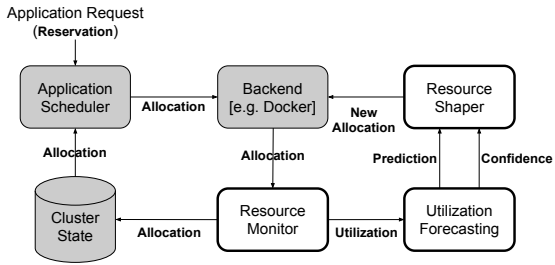

Figure 1: System overview: shaded boxes represent existing components, white boxes indicate new components presented in this work.

*What sets apart our approach from previous work is as follows. We use on-line forecasting with quantification of uncertainty to steer system behavior. This is necessary because, contrary to previous works, we explicitly take into account finite resources which, if handled improperly, can lead to failures. Additionally, we operate on low-level UNIX processes, and take control over the OS for shaping the resources allocated to applications.*

## 3 System Design

Figure 1 illustrates the architecture we assume in our work. The *backend* module is an instance of a cluster management system, such as Docker [18] or Kubernetes [26]. Additionally, we assume the presence of an *application scheduler* such as [42], which reads the compute cluster state from a dedicated database component. Finally, the monitoring component populates the cluster state database with measurements taken from the backend. In this Section, we focus on the two additional components we present in this paper: the *utilization forecasting* module, and the *resource shaper* module.

A bird's view on the operation of our system is as follows. Application execution requests take the form of resource reservations, which are submitted to the application scheduler. The application scheduler admits the request based on reservation information alone, and instructs the back-end to provision the necessary resources. The resource monitor collects information about both allocated and used resources, which are fed to the system state and the forecasting component respectively. The resource shaper module gauges resource allocation to match predicted utilization patterns, and is responsible for the preemption of running applications in case of sudden peaks in resource demand. The modified resource allocation is reflected in the system state, which in turn triggers new scheduling decisions. Next, we describe in detail the components that materialize our ideas.

**Resource monitor.** This module collects information about resource allocation and utilization from *every* component of *every* running application. This happens at regular time intervals: higher frequencies provide more accurate views, but generate more data. Our goal is to minimize intrusiveness by being application agnostic: for this reason we do not instrument applications (as done for example in [35]), but take standard metrics (CPU, memory, etc) as they are seen by the OS.

**Utilization forecasting.** The goal of this module is to anticipate the resource utilization of every application component. We study both parametric and non-parametric modeling approaches to predict resource utilization, with emphasis on the quantification of the uncertainty associated to these predictions. A more detailed exposition of the methodology we employ can be found in Section 3.1.

**Resource shaper.** This module uses utilization forecasts to adjust the resources allocated to every component of running applications. We anticipate prediction errors, thus we compensate using a "safe-guard" buffer of size $\beta$ to artificially increase (that is, to force over estimation) predicted peak resource utilization. A more detailed exposition of $\beta$ can be found in Section 3.2.

Additionally, the resource shaper is in charge of application preemption. Preemption policies can either be optimistic [54, 62] or strict (pessimistic). We advocate for a strict policy, to avoid delegating application preemption to the OS, which manages resource shortage (such as OOM) in an application agnostic and "unpredictable" way. A detailed exposition of the preemption policy can be found in Section 3.2.

### 3.1 Utilization Forecasting Module

The forecasting module is responsible for making predictions about future resource utilization, for each application component. For a given application, we forecast *both* CPU and memory utilization using monitoring data, which is available in the form of a time series that reflects resource usage across time[4]. We seek to discover patterns of resource usage that allow reasoning about our expectations on the future state of the system utilization.

We advocate for the need to **quantify the level of uncertainty** associated with each prediction: predictive errors may have serious impact on "finite" resources (i.e.

---
[4]Other types of resource can be considered as well.

memory), as they can cause application failures. Although errors are unavoidable to a certain extent, predictive confidence can be used to adjust the degree of adaptiveness to the anticipated workload: intuitively, a prediction with low confidence implies that the resource shaper should be conservative regarding changes in resource allocation.

In this work we compare the traditional *parametric* Autoregressive Integrated Moving Average (ARIMA) model to an alternative *non-parametric* model that offers a principled quantification of uncertainty. On the one hand, we use state-of-the-art ARIMA implementations that automatically tune hyper parameters and that provide a method to compute confidence levels associated to predicted values [8]. On the other hand, we model resource utilization using Gaussian Process (GP) regression [50], which is a Bayesian non-parametric regression method with many attractive features. Bayesian approaches control model complexity and thus avoid problems such as over-fitting [39]. Moreover, GPs offer a sensible framework for tuning their hyper parameters, through evidence maximization, that does not require cross-validation approaches which are typically more expensive and unpractical in the context of our work. Finally, the output of a GP regression model is a predictive distribution, rather than a single prediction, which allows reasoning about uncertainty in a principled way.

### 3.1.1 Time-series Prediction with ARIMA

ARIMA is often considered as the "go-to method" for time series forecasting: it is a generalization of the Autoregressive Moving Average (ARMA) model to cope with non-stationary time series data, which appear frequently in real-life applications such as the one we consider in this paper. Considering observation $y_t$ at time $t$, the ARMA($p'$,$q$) model is described as follows:

$$y_t - \alpha_1 y_{t-1} - ... - \alpha_{p'} y_{t-p'} = \epsilon_t + \theta_1 \epsilon_{t-1} + ... + \theta_1 \epsilon_{t-q} \quad (1)$$

where $\alpha$ are the parameters of the autoregressive part of the model, the $\theta$ are the parameters of the moving average part and the $\epsilon$ are error terms. In particular, $p'$ and $q$ are integers greater than or equal to zero and refers to the order of the autoregressive and moving average parts of the model respectively.

The underlying idea of ARIMA is that current values of a time series can be obtained by a linear combination of its past values, using finite differencing to produce stationary data. Formally, the ARIMA($p$,$d$,$q$) model using lag polynomials is given below:

$$(1 - \sum_{i=1}^{p} \phi_i L^i)(1-L)^d y_t = \delta + (1 + \sum_{I=1}^{q} \theta_i L^i)\epsilon_t \quad (2)$$

where $p = p' - d$, $\delta$ is a constant and $L$ is defined as the lag or back-shift operator. $d$ is an integer greater than or equal to zero and refer to the order of the integrated parts of the model and controls the level of differencing. Generally $d = 1$ is enough in most cases. An in-depth discussion about ARIMA can be found in [9].

In this work, model selection, that is, searching through combinations of order parameters to pick the set that optimizes model fit criteria, is carried out using the Akaike information criteria, a method that is widely available in most ARIMA implementations. Note that parameter optimization is an operation that needs to be performed multiple times during a forecasting period, to adapt to variations in the time series characteristics.

Finally, most ARIMA implementations output confidence intervals associated with the selected model parameters [9]. We note that confidence intervals should not be confused with prediction intervals: the former are associated to the probability of the true model parameters to be within the confidence interval, whereas the latter are associated to the likely range of future values output by the model. As discussed in the literature [9], confidence intervals for the mean are generally much narrower than prediction intervals. This has a direct consequence in the context of our work, which revolves around the idea of using predictive confidence to steer system behavior: for this reason, in the next section, we develop a Bayesian approach to time series modeling that features a principled approach to compute predictive confidence.

### 3.1.2 Time-series Prediction with GPs

In the GP literature, time series are treated as state space models, which are generalizations of auto-regressive models [41, 22]. Considering state $x_t$ and observation $y_t$ at time $t$, a state space model is described as follows:

$$\begin{aligned} x_{t+1} &= f(x_t) + \epsilon_t \\ y_t &= g(x_t) + v_t \end{aligned} \quad (3)$$

where $f(x_t)$ is the state transition function and $\epsilon_t$ is the process noise, which follows a normal distribution. The state $x_t$ may not be observed directly; an observation $y_t$ is given as a function of the state $g(x_t)$, which is additionally corrupted by observation noise $v_t$.

According to Equation (3), a time series is modeled as a non-linear Markovian dynamical system. The Markov property implies that the current state $x_t$ is *conditionally* independent from past states $\{x_\tau : \tau < t-1\}$, given the previous state $x_{t-1}$. The same is not true for the observations however. Thus, given a collection of noisy observations $\{y_\tau : \tau \leq t\}$, the goal for time series prediction is to infer the future state $x_{t+1}$. This requires learning the functions $f$ and $g$, which involves placing a GP prior over $f$ and $g$. However, the posterior over a non-linear dynamical system is not Gaussian, thus several approximation methods have been proposed in the literature [60, 21, 59].

In the context of recording resource utilization, we can make some simplifying assumptions. It is reasonable to assume that an observation $y_t$ matches the state $x_t$. Of course, we have to acknowledge that resource utilization constantly fluctuates; these fluctuations however can be sufficiently explained by the noise term $\epsilon_t$, which now accounts for both the process and the observation noise. We shall additionally make the dependency on past states explicit; for a history window of size $h$, we consider the following state-space model:

$$y_t = f(y_{t-1}, \ldots, y_{t-h}) + \epsilon_t \qquad (4)$$

To make predictions, we shall learn the transition function $f$ by means of standard GP regression. From Equation (4), the transition function depends on the history explicitly. In this way, we avoid the additional costs of approximating the true posterior of a non-linear dynamical system.

A GP model transfers information across points that are considered similar, as this is reflected in the choice of kernel $k(x, x')$, which determines the prior covariance between inputs $x$ and $x'$. If we assume that the inputs $\mathbf{X}$ solely consist of the recorded times, then similarity is only a matter of temporal locality, which is not optimal practice if the aim is to predict sudden changes of behavior throughout the course of a time series.

Hence, we resort to the definition of a kernel that relies on the observation history. It is implicitly assumed that if two sequences of observations are similar, then they must have been caused by the same "hidden" background processes; it is reasonable then to extrapolate and predict that the future observations will be similar as well. Such a history-dependent kernel can be easily constructed by transforming the data in an appropriate way. Consider a history window of size $h$, the training instances will be utilization patterns expressed as vectors of the form:

$$\tilde{\mathbf{x}}_t = [x_t, y_{t-h}, \ldots, y_{t-1}]^\top \qquad (5)$$

where $x_t$ is the $t$-th recorded time. Therefore, the history-dependent kernel is implemented by applying a typical exponential kernel on the transformed inputs:

$$k_h(x, x') = k(\tilde{\mathbf{x}}, \tilde{\mathbf{x}}') \qquad (6)$$

Two different inputs $x$ and $x'$ will be similar if they have a similar history pattern, or equivalently, if the $h$ preceded inputs have similar outputs. Note that we have kept the recorded times $x_t$ along with the history, thus we do not completely ignore locality in the original input space.

### 3.1.3 How Online Forecasting Works

From a practical perspective, the *forecast component operates in an online manner*. As long as new data is available, the predictive model will be trained and subsequently queried about the future workload. Depending on the modeling methodology, our approach is as follows.

**Using the ARIMA model.** The online training and prediction process that uses ARIMA operates by appending the new resource utilization data to the collection of observations gathered so far. ARIMA hyper-parameters are optimized using well-known methods [46, 51], which are known to be computationally expensive. Alternatively, works like [32] propose a stepwise algorithm (instead of using grid-search) that improves performance.

The $k$-step ahead forecast error is a linear combination of the future errors entering the system after time $t$:

$$e_t(k) = y_{t+k} - \hat{y}_t(k)$$

where $\hat{y}_t(k)$ is the estimated value. Since $\mathrm{E}[e_t(k)|y_t] = 0$, the forecast $\hat{y}_t(k)$ is unbiased with Mean Squared Error (MSE):

$$\mathrm{MSE}[y_t(k)] = \mathrm{Var}[e_t(k)]$$

Given these results, if the process is normal, the $100(1 - \alpha)$ forecast interval is:

$$[\, y_t(k) \pm N_{\alpha/2} \sqrt{\mathrm{Var}[e_t(k)]}\,]$$

where $N_{\alpha/2}$ is the multiplicative factor to obtain the percentile.

**Using the GP model.** The online training and prediction process that uses GP regression operates as follows:

1. New resource utilization data is appended to the collection of observations $\mathbf{X}, \mathbf{y}$. The rows of $\mathbf{X}$ are patterns as defined in Equation (5).

2. Using a history-dependent kernel $k_h(x, x')$, Equations (7) and (8) are used to make predictions based on observations $\mathbf{X}, \mathbf{y}$.

Under the assumption of a zero-mean prior and a Gaussian *likelihood*, that is, for any input-output pair we have $y \sim N(f(x), \sigma^2)$, the posterior is also a GP whose mean and covariance can be calculated analytically as follows:

$$\mathrm{E}[f(x) \mid \mathbf{X}] = k_h(x, \mathbf{X})(k_h(\mathbf{X}, \mathbf{X}) + \sigma^2)^{-1} \mathbf{y} \qquad (7)$$

$$\begin{aligned}\mathrm{Var}[f(x) \mid \mathbf{X}] = &\, k_h(x, x') \\ &- k_h(x, \mathbf{X})(k_h(\mathbf{X}, \mathbf{X}) + \sigma^2)^{-1} k_h(\mathbf{X}, x)\end{aligned} \qquad (8)$$

The predicted value at a new point will be the expectation under the posterior distribution, and the posterior variance quantifies the uncertainty about the prediction.

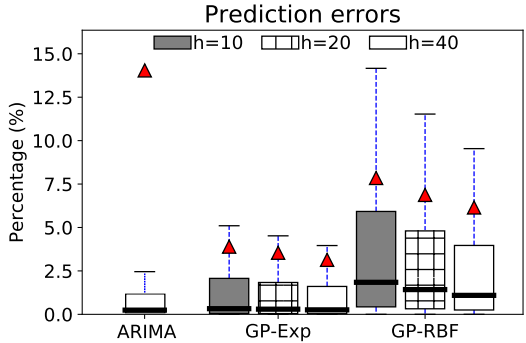

Figure 2: Boxplot showing error distribution of predicted utilization for a collection of time series in our academic cluster with different history points and, in case of GP, different kernels. The red triangle is the mean.

The regression step can be computationally expensive. Equations (7) and (8) involve a matrix inversion (for $k(\mathbf{X}, \mathbf{X}) + \sigma^2$), which is an operation of cubic complexity. Moreover, the set of observations $\mathbf{X}, \mathbf{y}$ will grow indefinitely during the lifetime of the system. While there is a plethora of methodologies on sparse GPs in the literature [58, 47, 48, 10], that can be used to reduce the complexity of regression, in this work we adopt the simple solution of restricting the dataset $\mathbf{X}, \mathbf{y}$ to the $N$ latest observations, thus keeping the model tractable. Note that $N$ is the number of patterns used; it should not be confused with $h$, which is the size of each pattern.

**Numerical results.** We have applied our modeling approaches on a dataset consisting of approximately 6000 time series that monitor the memory usage of applications in our academic cluster. Figure 2 summarizes the empirical distribution function for the predictive errors observed across the entire dataset, using ARIMA and GP.

In case of GP we forecast the future value using different number of past observations $h = [10, 20, 40]$, with $N = h$. As seen in Figure 2, increasing the value of $h$ results in smaller prediction errors. Also for the implementation of the history-dependent kernel as described in Equation (6), we have experimented both with the exponential and the squared-exponential (also known as RBF in the literature) functions. Figure 2 implies that the exponential implementation (GP-Exp) outperforms the RBF (GP-RBF) choice in terms of prediction error. Results for the GP are in line with our expectations, as the time series in question are typically not smooth. For the experiments of Section 4 and Section 5.1, we consider the exponential implementation of the history-dependent kernel only.

With ARIMA we observe that setting $p = h$ (so the autoregressive order equal to the history size) is overridden by hyper-parameter optimization, which yields $p \leq 3$.

Hence, the results for ARIMA do not depend on $h$. From Figure 2, it appears that ARIMA performs slightly better compared to GP for the median test error. Also the variance of the predictive error is smaller than with the GP model, an indication of a possible "over-confidence" in the model predictions. Our experimental results discussed in Section Section 4 corroborate this intuition: over-confidence leads to higher application failure rates, and an overall lower system efficiency, when compared to the GP model we present in this work.

### 3.2 Resource Shaper Module

We now delve into the details of the resource shaper module, which we use to adjust resource allocated to an application and its components as a function of predicted utilization. When resource are underutilized, the resource shaper "redeems" the excess capacity such that the application scheduler can dequeue idle applications. On the contrary, upon a utilization spike, the resource shaper needs to redeem resources from running applications and dedicate them to those experiencing a peak demand, for otherwise such applications are doomed to fail. Thus, the goal of the *preemption policy* we associate to the resource shaper is to decide how to redistribute resources, by operating on running applications and their components. Such a policy can optionally account for application priorities, as dictated by the application scheduler. Note that, irrespectively of the chosen preemption policy, *a failed application is resubmitted to the application scheduler*, making sure it enters the scheduling queue in a position commensurate to its original priority.

Recent works (for example [62]) advocate for an *optimistic* preemption policy, which is reminiscent of optimistic concurrency control [54]: resources are redeemed without taking explicit actions to manage the consequences of resource redistribution. Either explicit (and often manually set) priorities determine the fate of running applications, or the task is left to the OS.

Here, we present an alternative preemption policy, which we call *pessimistic*. Our goal is to control which application should be partially or fully preempted[5], while minimizing the amount of work that is wasted.

Algorithm 1 presents the details of our pessimistic preemption policy implemented by the resource shaper, which is triggered at regular time intervals, as determined by the output produced by the forecasting module. Given the current cluster state, and the resource utilization forecasts, the algorithm computes a new resource allocation

---

[5]We consider preemption primitives such as a `kill` operation, which inevitably waste work. Component or application suspension [45] and migration are outside the scope of this work. Alternatively, it would be interesting to consider techniques such as [29], which would allow a graceful management of memory pressure.

**Algorithm 1:** Overview of the pessimistic preemption policy implemented by the resource shaper module.

**Data:** $\mathcal{H} \leftarrow$ Hosts, $\mathcal{A} \leftarrow$ Running Applications

1   $cpusFree \leftarrow Array(\mathcal{H})$
2   $memFree \leftarrow Array(\mathcal{H})$
3   **foreach** $host \in \mathcal{H}$ **do**
4     $cpusFree[host] \leftarrow host.totalCpus$
5     $memFree[host] \leftarrow host.totalMem$
6   $\mathcal{J} \leftarrow \text{SORT}(schedulingPolicy, \mathcal{A})$
7   **foreach** $req \in \mathcal{J}$ **do**
8     $cpus \leftarrow cpusFree$
9     $mem \leftarrow memFree$
10   $remove \leftarrow False$
11   **foreach** $c \in req.CoreCpts$ **do**
12     $cpus[c.host] \leftarrow cpus[c.host] - c.futureCpus - \beta$
13     **if** $cpus[c.host] < 0$ **then**
14       $remove \leftarrow True$
15       **break**
16     $mem[c.host] \leftarrow mem[c.host] - c.futureMem - \beta$
17     **if** $mem[c.host] < 0$ **then**
18       $remove \leftarrow True$
19       **break**
20   **if** $remove$ **then**
21     $\text{INSERT}(req, \mathcal{K})$
22   **else**
23     $cpusFree \leftarrow cpus$
24     $memFree \leftarrow mem$
25     $E \leftarrow \text{SORT}(timeAlive, req.ElasticCpts)$
26     **foreach** $e \in E$ **do**
27       $cpus \leftarrow cpusFree[e.host] - e.futureCpus - \beta$
28       $mem \leftarrow memFree[e.host] - e.futureMem - \beta$
29       **if** $cpus \leq 0$ **or** $mem \leq 0$ **then**
30         $\text{INSERT}(e, \mathcal{K_E})$
31       **else**
32         $cpusFree[r.host] \leftarrow cpus$
33         $memFree[r.host] \leftarrow mem$
34   **foreach** $req \in \mathcal{K}$ **do**
35     **foreach** $c \in (req.CoreCpts \cup req.ElasticCpts)$ **do**
36       $\text{PREEMPCOMPONENT}(c)$
37   **foreach** $e \in \mathcal{K_E}$ **do**
38     $\text{PREEMPCOMPONENT}(e)$
39   **foreach** $req \in \mathcal{J} \setminus \mathcal{K}$ **do**
40     **foreach** $c \in (req.CoreCpts \cup req.ElasticCpts)$ **do**
41       $\text{RESIZECOMPONENT}(c)$

for each running application, which is then imposed on the cluster by operating directly on application components through low-level preemption primitives.

The algorithm starts by initializing (lines 1-5) the variables that holds the information about the allocated resources. Then it sorts (line 6) running applications according to the application scheduler policy (e.g.; FIFO, that is, arrival times), and it computes (lines 7-33) an allocation by trying to maximize the resource allocation while minimizing the number of running applications. In particular, it first allocates the core (lines 8-19) components and then all elastic components[6] that fit in the host (lines 23-33). The algorithm continues until all running applications are processed.

Resource allocation is determined, and we can turn our attention to preemption. Core components that no longer fit a host entail full application preemption (lines 34-36). Also elastic components can be preempted (lines 37-38), inducing only a partial application preemption. In addition, in case of elastics components, we can experience partial or entire loss of the work done by the preempted component. For this reason, our algorithm allocates the core components of an application, then moves to the elastic components by giving priority to the ones that have been living in the cluster for a longer time (line 25). Components recently scheduled are the best candidates for preemption, because they have likely produced less useful work. Finally, the algorithm resizes (lines 39-41) the components according to the computed allocations. Our algorithm currently supports CPU and Memory, but it can be extended to other types of resource as well.

**Safe-guard buffer.** We are now ready to define the "safe-guard" buffer. The buffer size $\beta$ is a function of the uncertainty quantified by the forecasting module:

$$\beta = K_1 R_{A_i} + K_2 V_{A_i} \qquad (9)$$

where $R_{A_i}$ is the initial resource **request** for application $A_i$, and $V_{A_i}$ is the estimated variance of the prediction, as these are given by the forecasting module (ARIMA or GP). Equation (9) involves a constant term $K_1 R_{A_i}$ and a dynamic term $K_2 V_{A_i}$. The constant term can be though of as a *minimum resource allocation* that is granted to application $A_i$. The dynamic term uses the confidence (expressed as variance $V_{A_i}$) given by the predictor to adjust $\beta$ accordingly: it thus changes during an application lifetime. In Section 4, we study how different values of $K_1$ and $K_2$ affect the performance of our method.

## 4 Simulation-based Evaluation

### 4.1 Methodology

We evaluate our mechanism using an event-based, trace-driven discrete simulator which was developed to study the scheduler Omega [54], and was later extended in [42] to study application schedulers. We have made additional extensions[7] to support the concepts of this work.

---
[6] In case the application scheduler does not support the distinction between core and elastic, all components are treated as core.

We use publicly available traces [63, 52, 53, 25], and generate a workload by sampling from the empirical distributions computed from such traces. Our workload is composed by 150.000 batch applications, both *rigid* (e.g. TensorFlow) and *elastic* (e.g. Apache Spark) variants. Applications are assigned a number of components ranging from a few to tens of thousands. The resource requirements of application components follow that of the input traces, ranging from a few MB of memory to a few dozens of GB, and up to 6 CPU cores. Application *runtime* is generated according to the input traces, and ranges from a few dozens of seconds to several weeks (of simulated time). Inter-arrival times are drawn from the empirical distributions of the input traces, and exhibit a bi-modal distribution with fast-paced bursts, and longer intervals between application submissions.

We simulate a cluster consisting of 250 homogeneous machines, each with 32 cores and 128GB of memory. All results shown here include 10 simulation runs, for a total of roughly 3 months of simulation time for each run.

The metrics we use to analyze the results include: *(i)* application **turnaround**, which allows reasoning about the scheduling objective function, *(ii)* **resource slack**, measured as the difference of percentage of CPU and memory the scheduler allocates to each application compared to the percentage actually used by the application and *(iii)* application **failures**, which give us information about the aggressiveness of our approach.

## 4.2 Results

Next, we present experimental results that demonstrate the advantage of our resource shaping mechanism, compared to a baseline approach which matches allocation to reservation. Two alternatives for time series prediction are examined. We first consider an ideal setup with an oracle having perfect information about future workload: this allows to determine an upper bound of the performance gains achieved by our approach. Then, we compare the two models developed in Section 3.1 (ARIMA and GP), to investigate the impact of prediction errors on system performance.

**Baseline.** It constitutes a reservation centric approach (similar to Mesos and Yarn, as originally implemented in the Omega simulator [54]) that achieves the performance reported in Figure 3. This approach relies entirely on the resource requested by the application (when submitted) in order to allocate resources in the cluster and does not modify them at runtime.

[7]https://github.com/DistributedSystemsGroup/cluster-scheduler-simulator

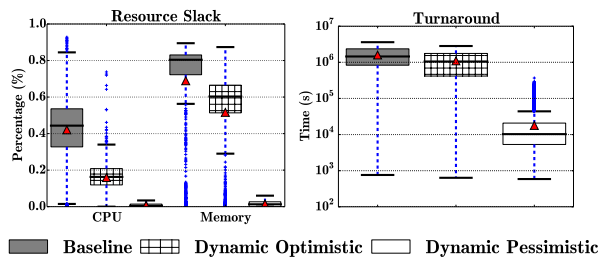

Figure 3: Boxplots comparing baseline vs optimistic vs pessimistic approaches over different metrics, using an oracle in place of the prediction module. The red triangle is the mean.

**Oracle-based resource shaping.** We gloss over prediction errors induced by a real statistical model and consider an ideal scenario from the forecasting point of view. Ultimately, our goal is to discern virtues and drawbacks of different preemption policies. Results are summarized in Figure 3: the plots correspond to resources slack and application turnaround, whereas each box correspond to the baseline and our resource shaping approach, with an optimistic (as originally implemented in the Omega simulator [54]) and our pessimistic preemption policy. Note that our simulator implements the concept of work lost when an application component crashes or gets killed.

Overall our results indicate that resource shaping brings substantial benefits in terms of all metrics we consider, in the absence of prediction errors. Cluster efficiency improves because resource slack, computed as the difference between allocated and used resources, drastically shrinks as shown in Figure 3 (left) compared to the baseline. Similarly, turnaround times are notably smaller as shown in Figure 3 (right) in comparison to the baseline. Indeed, the system can ingest new applications more quickly, because resources are better used.

Figure 3 can now be used to compare optimistic versus pessimistic eviction policies, in absence of prediction errors. While both approaches improve over the baseline, the pessimistic policy we introduce in this work is consistently superior to the optimistic policy in all respects. As shown in Figure 3 (left), the pessimistic policy induces our resource shaping mechanism to follow very closely application resource utilization: in this case, resource slack becomes negligibly small. This result explains why turnaround times, Figure 3 (right), are almost two orders of magnitude smaller with the pessimistic policy: by freeing up resources, the application scheduler is amened to trigger new executions, thus queuing times shirk. Furthermore, we compute the number of application failures: in case of the optimistic policy we record 37.67% application failures, whereas with the pessimistic policy no application fails. Indeed, with the optimistic policy, when two applications compete for resources and there are none left, the system will let one of the two fail. Instead, the pessimistic policy avoids failures through partial preemption, by freeing elastic resources first.

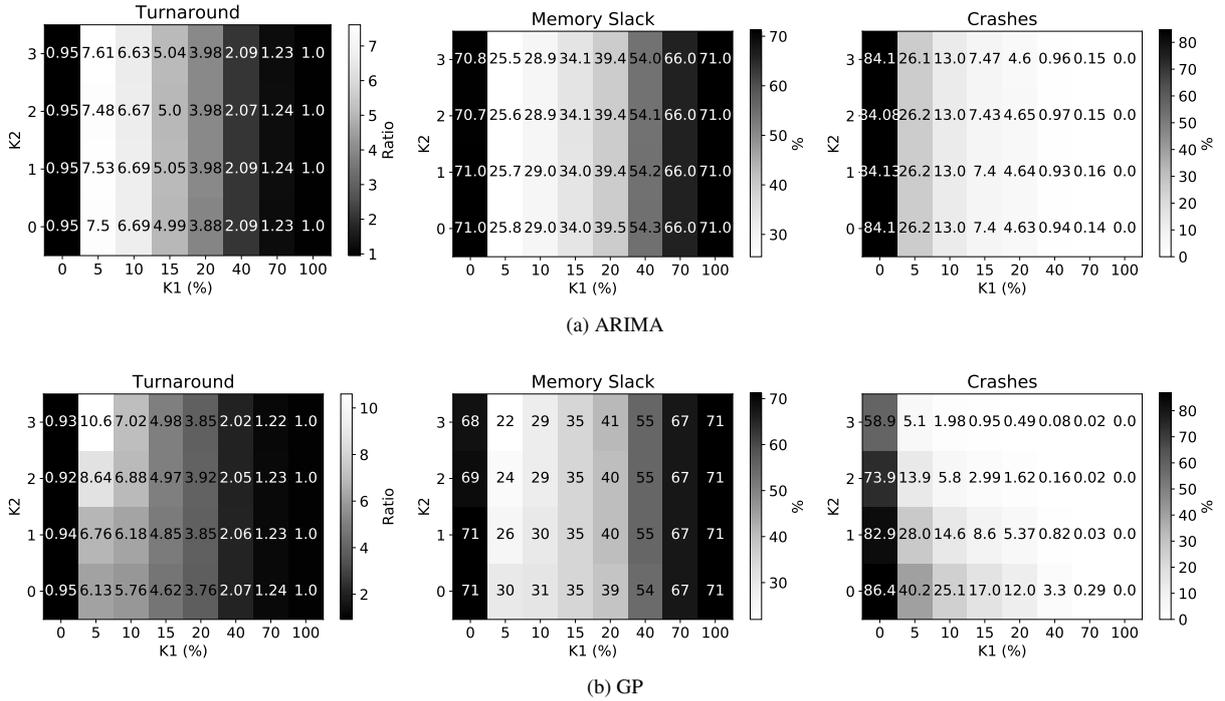

Figure 4: Heat maps showing the effect of $K_1$ and $K_2$, which compose $\beta$, on different metrics when using ARIMA and GP. Bright cells are better.

**ARIMA-based resource shaping.** Next, we study the system behavior when using ARIMA to predict future resource utilization. As anticipated in Section 3, statistical models are prone to prediction errors, which we address using the buffer $\beta$. A key feature of our approach is that $\beta$ is a function of the uncertainty produced by the model. In practice, when the predictor outputs a future (peak) resource utilization, we adjust the value by adding the buffer $\beta$. In Figure 4a we demonstrate the effect of the buffer parameters ($\beta = f(K_1, K_2)$) on the turnaround ratio over the baseline, the memory slack and application failures (we show average results). In all cases, bright cells are better.

On the x-axis, $K_1$ controls the static component of Equation (9), which gauges the minimum amount of resources systematically granted to applications. The value of $K_1$ is expressed as a percentage of the requested resources; when $K_1 = 100\%$ our approach degenerates to the baseline. On the y-axis, $K_2$ controls the dynamic component of Equation (9), which integrates prediction uncertainty. We let $K_2$ vary in the range $[0, 1, 2, 3]$ which define different bands around the mean of the predictive Gaussian distribution, according to the three-sigma rule.

Let's first slice Figure 4a by row, and focus on the $K_2 = 0$ case: here we omit uncertainty information and only consider the effects of a static, minimum resource allocation. Even with just $K_1 = 5\%$, our approach achieves 7.5x average improvements in terms of application turnaround, while resource slack is only 30% in average. However, the number of crashed application is high: roughly 26% of applications experience a failure in average, and the situation improves only for large values of $K_1$. In the limit, when $K_1 = 100\%$, our method degenerates to the baseline: here no application fail, but turnaround times and slack exhibit no improvements. In our system, when an application crashes it is resubmitted and, after a certain amount of failures, the system is not shaping its allocation anymore. Also, even if applications crash they can still benefit from being able to start sooner than a baseline system because other applications were able to complete their work sooner.

We note that the absence of a static term (i.e. $K_1 = 0\%$) results in turnaround that is very close to the baseline regardless of $K_2$, due to the high number of applications failures which also lead to an high memory slack. This is a consequence of the occasional high confidence of the predictor in cases where a sudden change in the usage behavior occurs. It is necessary to maintain a static com-
10

ponent to accommodate unexpected variations, which are very difficult to capture with statistical methods.

Finally, we focus on $K_1 = 5\%$: the minimum resource allocation is small, and we absorb prediction errors and fluctuations using uncertainty information. However, as $K_2$ increases, all metrics remain similar: the uncertainty produced by the ARIMA model is not sufficiently accurate to compensate forecasting errors.

**GP-based resource shaping.** Next, we study the system behavior when using GP regression to predict future resource utilization. Similarly to the ARIMA-based resource shaping, in Figure 4b we demonstrate the effect of the buffer parameters. However, we can see that while GP gives slightly worst results when not considering the uncertainty of the forecasting values ($K_2 = 0$) compared to ARIMA, as $K_2$ increases, all metrics improve: average turnaround ratios increase up to 10.6x improvement, average slack is reduced to a 22% in average, while application failures quickly decrease.

In our setup, the best performance is achieved when the system is most flexible regarding the size of the buffer, i.e., a high value for its dynamic and a small value for its static components.

*In summary the results show that, for the best configuration of parameters with a real predictor and not an oracle, tunraround time and resource slack is more than halved in the median case, both in terms of CPU and memory resources. By using the uncertainty provided by the forecasting model based on the GP, we are able to improve these metrics further, achieving 10.6x improvement compared to the baseline for the turnaround time.*

## 5  System Implementation

We materialize the ideas presented in this paper with a full-fledged, python-based, implementation of our mechanism, following the system design presented in Section 3, and depicted in Figure 1. For this work, we build the resource shaper to interact with the application scheduler presented in [42], which we recently adopted to manage our workloads. In our implementation, the resource shaper modulates both CPU and memory resources.

In our cluster, we use Docker [17] as the back-end and we have investigated how to resize its containers (corresponding to application components). There are two values that Docker uses to check for Memory limits: a hard and a soft limit. When the hard limit is surpassed, the container is killed by the OS. Instead, when the soft limit is reached, the OS tries to release some resources first. In our work we use the soft limit value since the application scheduler we use takes decisions based on such value. In particular, we rely on the OS low level mechanisms to notify the processes running in the container to free some of their resources. This practice is compatible with frameworks such as the Java Garbage Collector (GC) that attempts to release allocated but unused memory space. Note that our technique is compatible with approaches such as [30], which trade performance for a smaller memory footprint.

The monitoring component feeds the utilization forecast module with data at regular time intervals. Frequent updates ultimately result in better system efficiency, as the predictor operates on a high-fidelity view of resource utilization in the cluster. However, this might impose a high toll in terms of monitoring scalability. On the other hand, infrequent updates improve scalability at the expense of lower system efficiency and responsiveness. In our implementation, we collect resource utilization information every minute, which is in line with what done in [62].

Next, we provide additional details of our prototype.

**Forecasting module.** It implements the two models we discuss in Section 3.1. For the ARIMA model we use the well-known `StatsModel` [55] library, which features an efficient implementation of the ARIMA model and its automatic parameter tuning through the Pyramid wrapper [24]. For the GP model we use the well-known library `GPy` [57]. Both models consider a small history of the ten past observations for training, to keep computational complexity under control.

**Resource shaping module.** It materializes the ideas presented in Section 3.2. The ultimate goal of the resource shaper is that of issuing commands to preempt (kill, in our implementation) an entire application, or individual components thereof, and to resize the resource allocation, as computed by the by Algorithm 1. It is important to point out that the resource shaper adapts resource allocations only after enough historical data points are available for the forecasting module: we call this a **grace period**, and set it to 10 minutes in our experiments.

The resource shaper uses the mechanisms exposed by Docker (as discussed above) to adjust application resources, and to eventually preempt components or entire applications. This module computes a new resource allocation for all running application in the system, based on the predicted value and variance obtained from the forecasting module. The buffer $\beta$ is set to compensate for prediction uncertainty, using the parameters that we obtain through simulations, that is $K_1 = 5\%$ and $K_2 = 3$.

### 5.1  Experimental Evaluation

We have deployed the mechanism presented in this paper in our cluster, which we operate using [20]. Our goal is to perform a comparative analysis between dynamic resource shaping and a baseline, as done in Section 4. The baseline system supports the concept of distributed appli-cations [42], but follows a reservation centric approach, in which allocation matches reservation for the entire application lifetime. In our experiments, we consider exactly the same workload trace on both systems which takes approximately 24 hours from the first submission to the completion of the last application.

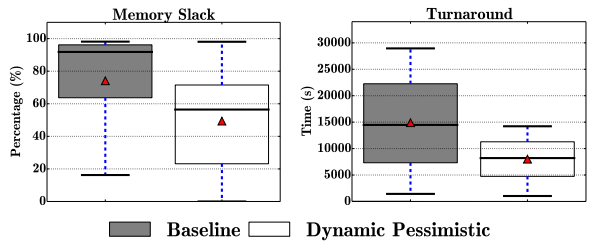

Figure 5: Boxplots comparing baseline vs pessimistic dynamic approach over memory slack and turnaround time distributions using GP-based resource shaping. The red triangle is the mean.

**Workload.** We use two representative *application templates* including: 1) an elastic application using the Apache Spark framework; 2) a rigid application using the TensorFlow framework. Similarly to the traces used in Section 4, we set our workload to include 60% of elastic and 40% of rigid applications, for a total of 100 applications. Application inter-arrival times follow a Gaussian distribution with parameters $\mu = 120$ sec, and $\sigma = 40$ sec, which is compatible with what we observe in our cluster. Regarding the elastic application templates, we consider three use cases. First we consider an application that induces a random-forest regression model to predict flight delays, using publicly available data from the US DoT [19]. Second we consider a music recommender system based on the alternating least squares algorithm, using publicly available data from Last.fm [36]. Third we consider an Extract, Transform and Load (ETL) application. All applications have 3 different flavors: while they all have 3 core components, the number of elastic components varies depending on the flavor. In terms of RAM, all flavors have different reservation values that span from 8GB to 32GB. Instead, using the rigid application template, we train a deep GP model [12], and use a single TensorFlow instance, with 1 worker and 8-16-32GB of RAM depending on the flavor.

**Experimental setup.** We run our experiment on a isolated platform (which we use as testbed for non-production systems) with ten servers, each with a 8-core CPU running at 2.40GHz, 64GB of memory, 1Gbps Ethernet network fabric and two 1TB hard drives each. The servers use Ubuntu 14.04 and Docker 17.09.0. Docker images for the applications are preloaded on each machine to prevent startup delays and network congestion.

**Summary of results.** Using the FIFO scheduling policy, and the GP-based utilization forecasting module, we compare the two systems, baseline and dynamic. Overall, the dynamic system is largely more efficient and responsive. We measure substantial improvements in terms of resource allocation: indeed our system can afford to ingest more applications, that would otherwise wait to be served. Figure 5 (left) illustrates resource slack, which is roughly 40% lower with our resource shaping mechanism. As a consequence, applications spend less time in the scheduler queue and have short turnaround times, as shown in Figure 5 (right). The median turnaround times are $\sim 50\%$ shorter. Note also that the tails of the distributions are in favor of our approach. Finally, we report that no application, nor component failed when using our resource shaping mechanism, configured with the pessimistic preemption policy.

# 6 Conclusions

The emergence of "the data-center as a computer" paradigm has led to unprecedented advances in cluster management frameworks, that aim at exposing distributed, cluster resources to a variety of business-critical and scientific applications. However, the current resource reservation model hinders an efficient use of cluster resources. Resource utilization dynamics induce over-provisioning, which is one of the main culprit of poor efficiency. The problem of underutilization has been addressed by several approaches. For example, the design of economic incentives to steer system operation has led to the development of complex resource markets, e.g. AWS Spot instances, which call for the design failure tolerant applications, due to the ephemeral nature of the resources they are offered.

In this work, we presented a mechanism that cooperates with a scheduler to dynamically adjust resources allocated to an application, so that they closely match those they actually use throughout their lifecycle. Our design featured: a method to build a statistical model to forecast resource utilization, and a preemption policy that reallocates system resources while minimizing failures.

We have validated our mechanism numerically and with a real experimental campaign. Our simulations shed lights on the key role played by our ability to model and use prediction uncertainty, and by the use of strict preemption vs. optimistic concurrency control. We implemented a system prototype of our dynamic allocation mechanism and deployed it in a test environment, where we executed a real workload. Results indicate notably improved system efficiency, which translates in better responsiveness.